\documentclass[prb,twocolumn,aps,showpacs]{revtex4}
\usepackage{graphicx}%
\usepackage{dcolumn}
\usepackage{amsmath}
\usepackage{amssymb}
\usepackage{epsfig}
\usepackage[latin1]{inputenc}
\usepackage{amsfonts}
\usepackage{caption}
\include{biblio}
\usepackage{color}
\usepackage{needspace}
\usepackage{makeidx}
\usepackage{appendix}
\begin{document}

\title{Spin-orbit-coupled  atomic Fermi gases  in  two-dimensional optical lattices in the presence of a Zeeman field}
\author{Zlatko Koinov, Shanna Pahl  }\affiliation{Department of Physics and Astronomy,
University of Texas at San Antonio, San Antonio, TX 78249, USA}
\email{Zlatko.Koinov@utsa.edu}
 \begin{abstract}
   We investigate the single-particle and collective excitations of a Rashba spin-orbit coupled atomic Fermi gas with attractive interaction, loaded in a two dimensional (2D) square optical lattice, in the presence of an effective out-of-plane  Zeeman field. Our numerical calculations show that the many body
physics of the Bardeen-Cooper-Schrieffer (BCS) side is strongly modified compared to  the Fermi gases in the free space. The physics behind this statement is in the fact that without a lattice structure, if the value of the Zeeman field does not exceed some threshold value, the minimum of the single-particle ground state energy is infinitely degenerate and occurs along a ring in $(k_x,k_y)$ space. This reduces the effective dimensionality; the  single-particle density of states is a constant at low energies, and the molecular pairing is strongly enhanced.  In contrast to the continuum, in the 2D lattice case there are only four degenerate minima, and because of that,  the spin-orbit coupling (SOC) in optical lattices gives rise to unusual properties entirely different from the   continuum. For example, in the continuum, the pairing gap as well as the condensate fraction are strongly enhanced by the SOC strength on the BCS side. In the presence of lattice geometry the gap and the condensate fraction increase as a function of the SOC strength only at the small fillings and weak attraction limit. Moreover, we found that the speed of sound  also exhibits different behavior: in the 3D and 2D continuum, the slope of the Goldstone mode decreases as a function of the SOC strength, while in the lattice case the speed of sound increases monotonically with SOC strength.
\end{abstract}
\maketitle
\section{Introduction}
The experimental realization of  spin-orbit coupling (SOC) in ultracold atoms \cite{Lin1,Lin2,Lin3} has  initiated substantial  theoretical efforts to study its effects in 2D and 3D Fermi gases \cite{SOC1,SOC2,SOC3,SOC4,SOC5,SOC5a,SOC6,SOC7,SOC8,SOC9,SOC10,SOC11,SOC12,SOC13,SOC14,SOC15,SOC16,SOC17,SOC18}. This is because SOC systems with a Zeeman field, which breaks the population balance, can lead to novel types of exotic superfluid phases that may support Majorana fermions. \cite{M1,M2}

 Although the Rashba SO coupled atomic Fermi gases in the free space have been extensively investigated, the entirely different case of attractive Fermi gas in a square optical lattice with SOC  has not been completely  studied.\cite{OL1,OL2,OL3,OL4,OL5,OL6,OL7,OL8} In what follows, we shall study a population-balanced mixture of fermion atoms with an s-wave pairing interaction loaded in a square two-dimensional  (2D) optical lattice, with Rashba SOC and an external out-of-plane  Zeeman field. We shall use a small Zeeman field, because in this regime the pairing between atoms on the same Fermi surface is preferred; this is the normal BCS superfluid (for a larger Zeeman field, the topological BCS superfluid as well as the Fulde-Ferrell-Larkin-Ovchinnikov superfluid phases appear).\cite{OL6}
\subsection{SO coupled atoms in continuous space}
  It is widely assumed that the two-body problem is the key to understanding the physics of SO coupled Fermi gases. Consider, for example, the two-body problem in the case of a three-dimensional (3D) spin-$1/2$ Fermi gas with  a Rashba SOC and an external out-of-plane  Zeeman field. The spectrum of the single-particle excitations is $\epsilon_{\pm}(\textbf{k})=\varepsilon(\textbf{k})\pm\sqrt{h^2+\lambda^2k^2_\perp}$, where $\varepsilon(\textbf{k})=\hbar^2\textbf{k}^2/(2m)-\mu$, $\mu$ is the chemical potential, $h$ is the strength of the Zeeman field and $\lambda$ is the SOC constant. A bound state exists if its energy $E_B<0$ is less than twice the minimum of the single-particle
ground state energy $\epsilon_{min}$, i.e. $E_B-2\epsilon_{min}=\varepsilon_b<0$. If the value of the Zeeman field exceeds the  threshold value $h>h_{c}=m\lambda^2/\hbar^2$, a unique lowest single-particle state $\epsilon_{min}=-h$ occurs at  $\textbf{k} = 0$, and the bound state  exists only on the BEC side of the Feshbach resonance. If $h<h_{c}$, the minimum single-particle energy $$\epsilon_{min}=-\left[m\lambda^2/(2\hbar^2)+\hbar^2h^2/(2m\lambda^2)\right]$$ occurs continuously along a ring  of radius  $k_\perp=\sqrt{m^2\lambda^2/\hbar^2-\hbar^2/\lambda^2}$ in the $(k_x,k_y)$ plane.\cite{SOC4} Since the single particle ground state
is not unique, the ground state of a boson condensate is also not unique, and because the density of states (DOS) is a
constant at low energies, a two-body bound state appears for any weak attractions, and therefore, the bound state exists even on the BCS side. The corresponding two-body bound state energy $E_B$ can be obtained by numerically solving the two-body bound state equation. In the Gaussian approximation, $E_B$ at  zero temperature  is determined by the following equation:\cite{SOC4,SOC6,SOC7,SOC11,SOC14}
\begin{equation}\begin{split}\label{2EB}
&\frac{1}{U}=-\frac{1}{2N}\sum_{\mathbf{k}}\left[\frac{1}{E_B-2\epsilon_+(\textbf{k})}+\frac{1}{E_B-2\epsilon_-(\textbf{k})}\right] \\ &+\frac{1}{N}\sum_{\mathbf{k}}\left[\frac{4h^2}{(E_B-2\varepsilon(\textbf{k}))(E_B-2\epsilon_+(\textbf{k}))(E_B-2\epsilon_-(\textbf{k}))}\right].
\end{split}\end{equation}
Here,  the  interaction strength $U$ needs to be regularized in a standard manner by means of the s-wave scattering length $a_s$.  According to the solutions of Eq. (\ref{2EB}),  an increase of the SOC strength $\lambda$
leads to a deeper bound state, i.e.  the binding energy strongly depends on the SOC strength. Since  these bound states are caused by the SOC, they have been referred to as rashbons.\cite{SOC4,SOC6,SOC7,SOC11,SOC14}

In the many-particle problem, the single-particle spectrum is $E_{\pm}(\textbf{k})=\{\lambda^2k^2_\perp+\Delta^2+h^2+\varepsilon^2(\textbf{k}) \pm 2\sqrt{h^2\Delta^2+
\varepsilon^2(\textbf{k})\left[h^2+\lambda^2k^2_\perp\right]}\}^{1/2},$
 where the chemical potential $\mu$ and the pairing gap $\Delta$ can be obtained  by numerically solving  the corresponding number
and gap equations. It was determined, in Ref. [\onlinecite{SOC6}], that in the large SOC limit the chemical potential $|\mu|=|E_B|/2-\mu_b$ becomes negative since $\mu_b$, the chemical potential for composite bosons (which is positive due to the repulsion between rashbons), decreases with increasing $\lambda$, as shown in the upper inset of Fig. \ref{F4}.

The convergence of the chemical potential and one-half of the bound state energy on the BCS side, evident in the two-body
problem, suggests that the Rashba SOC may trigger a new type
of crossover to rashbon BEC  in the many-body problem of fermions. Thus, the BCS-BEC crossover is of special interest, and it has been stated that the contributions of the singlet and triplet pairings to the condensate fraction, separately, characterize the crossover better than other quantities. \cite{SOC5}
\begin{figure}[htf]
	\centering\includegraphics[scale=0.25]{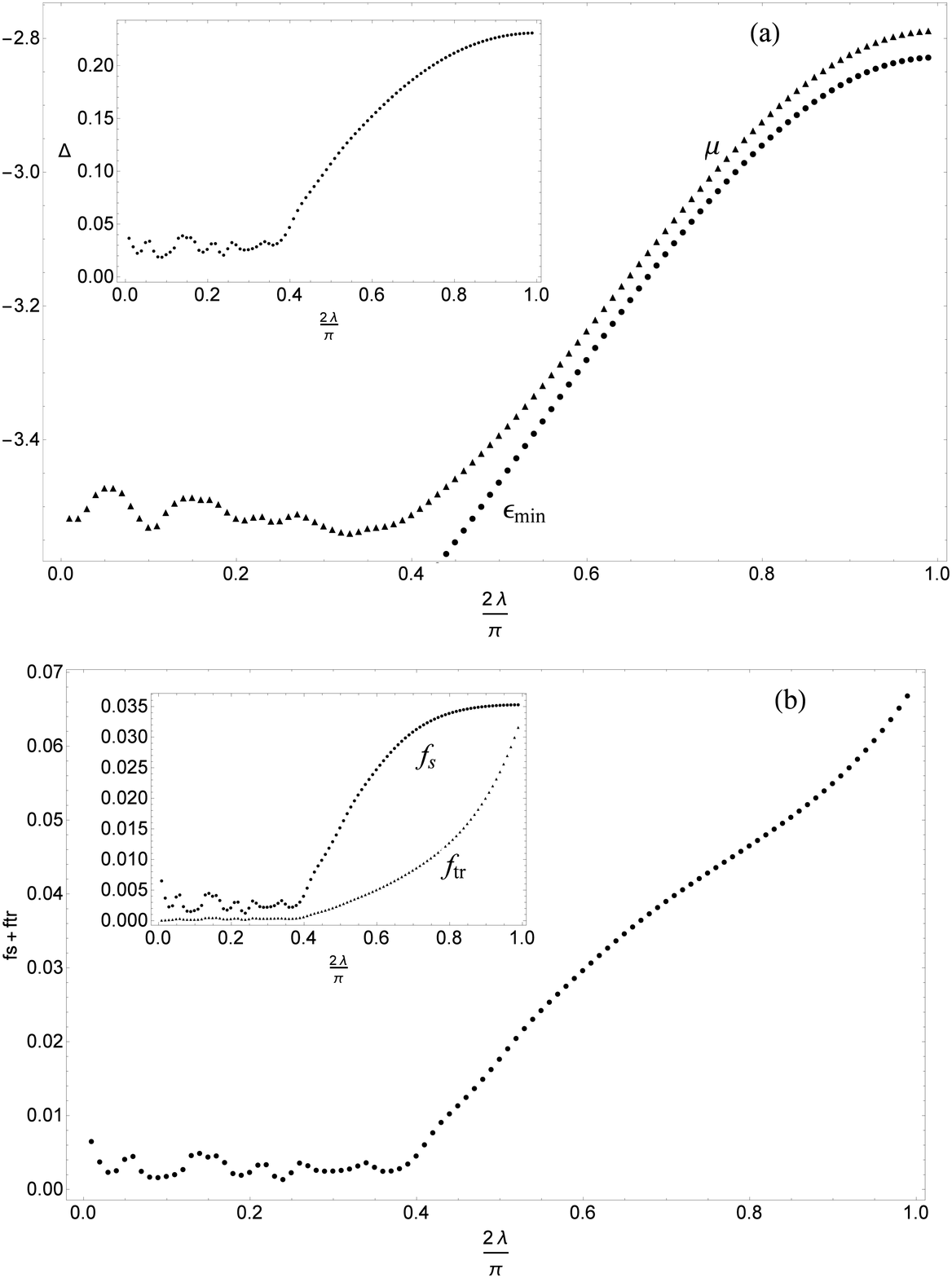}
	\caption{
Chemical potential $\mu$, pairing gap $\Delta$, the minimum of the single-particle ground state energy $\epsilon_{min}$ (a), and (b) the total condensate fraction $f_c=f_s+f_{tr}$ ($f_s$ and $f_{tr}$ are the singlet and the triplet contributions) of a Fermi gas in a square optical lattice subject to a Rashba non-Abelian SOC of strength $\lambda$. The on-site attractive strength is $U=2t$, and the filling factor is $f=0.1$. }\label{F1}
\end{figure}
\begin{figure}[htf]
	\centering\includegraphics[scale=0.5]{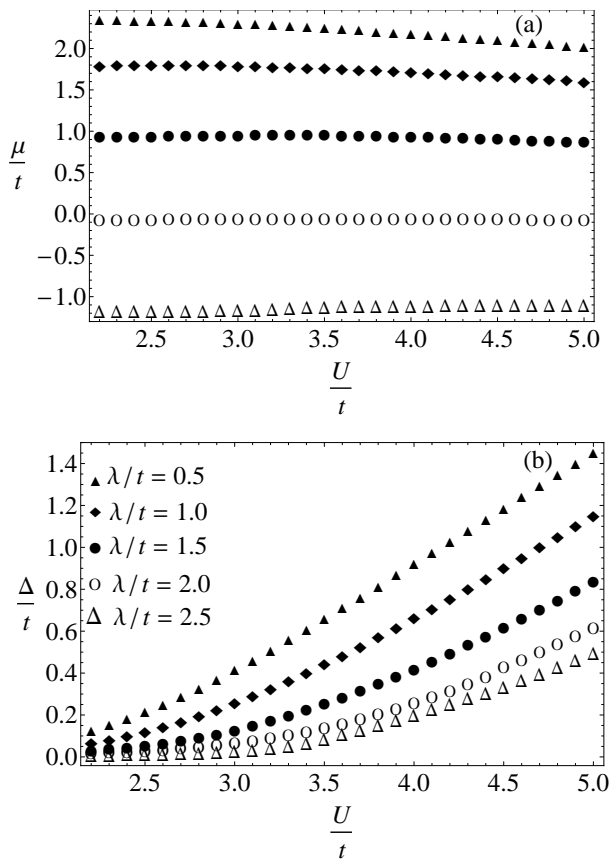}
	\caption{Zero temperature mean-field chemical potential $\mu$ (a) and gap $\Delta$ (b) for Fermion atoms in a 2D optical lattice as a function of the strength of the attractive interaction $U$ for different values of the  Rashba SOC strengths $\lambda$ at a Zeeman field $h=0.4 t$ and $f=0.5$.}\label{F2}
\end{figure}
\begin{figure}[htf]
	\centering\includegraphics[scale=0.4]{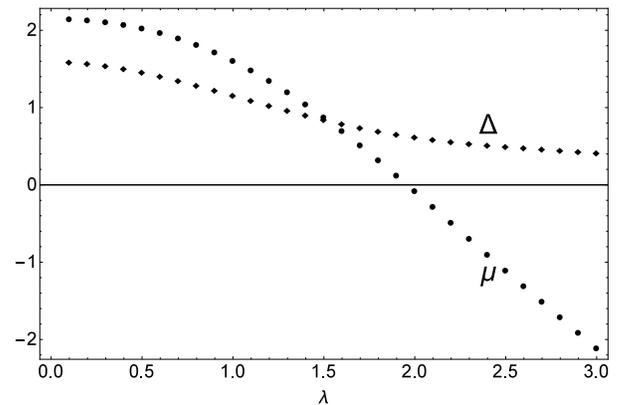}
	\caption{Zero temperature gap $\Delta$ and chemical potential $\mu$  as a function of the Rashba SOC strength $\lambda$ for Fermion atoms in a 2D optical lattice. The system parameters are $U=5t$, $f=0.5$ and $h=0.4 t$ ($\lambda$, $\Delta$ and $\mu$ are in units of $t$). }\label{F3}
\end{figure}

\subsection{Square optical lattice with an external non-Abelian gauge field }
The system of a two-component Fermi gas moving in an optical square lattice with an external non-Abelian gauge field has been considered in Refs. [\onlinecite{OL1,OL2}]. Let us assume that the spin-conserved hopping term is
proportional to $t\cos(\lambda)$, and the spin-flipped term is in proportion to $t\sin(\lambda)$  (the energy units are in terms of the tunneling strength $t=1$, and spatial units are in terms of lattice constant $a=1$). The lowest
single-particle state $\epsilon_{min}=-2\sqrt{3+\cos(2\lambda)}$ is four times degenerate, and occurs at four points in momentum space
$$\pm k_x , \pm k_y=\cos^{-1}\left(\frac{\cos(\lambda)}{\sqrt{\cos^2(\lambda)+\sin^2(\lambda)/2}}\right).$$

As pointed out in Ref. [\onlinecite{OL1}], hereafter referred to as NAL, the
interplay between interactions and the Dirac spectrum generated by the Rashba coupling, leads to interesting behavior with respect to the filling factor, especially in the weak and intermediate attraction regimes. The authors of NAL noted that for small fillings, not only does chemical potential increase with SOC (as opposed to the continuous system), the pairing gap and condensate fraction are enhanced by SOC. However, for close to half filling, they are suppressed.

To illustrate this, in Fig. \ref{F1} we have shown: (a) the chemical potential, gap, and $\epsilon_{min}$, and (b) the condensate fraction, all calculated by numerically solving the corresponding number and gap equation in the mean-field approximation for filling $f=0.1$ and weak interaction strength $U=2t$.   As can be seen, above critical SOC $\lambda_c=0.4\pi/2$, the chemical potential approaches $\epsilon_{min}$, and both the gap and the total condensate fraction increase rapidly. This is a fingerprint for the formation of SOC-induced bound states.  Furthermore, the authors of NAL show that $\lambda_c$ increases when the filling is increased, but after $f=0.7$, the condensate fraction begins to decrease.

 In what follows, we present our study of the interplay between SOC, the strength of the interaction,
 and the Zeeman field in lattice systems, and compare our findings with the aforementioned non-Abelian lattice system (NAL), and free space systems.

 The paper is organized as follows. In Sec. II, we use the mean-field single particle Green's function to obtain the singlet and the triplet pairing amplitudes, and the singlet and triplet condensate fractions. In Sec III, we use the functional-integral formalism to obtain the Bethe-Salpeter (BS) equation in the  generalized
random phase approximation (GRPA) for the spectrum of the two-particle excitations. In the GRPA, the single-particle excitations are
replaced with those obtained in the mean field approximation;
while the collective modes are obtained by solving
the BS equation, in which the single-particle Green's
functions are calculated in the mean field approximation,
and the BS kernel is a sum of the direct and exchange
interactions. In diagrammatic language, the kernel is
represented by ladder and bubble diagrams. We compare the speed of sound, obtained by the BS formalism, to the corresponding speed obtained in the Gaussian approximation. Since the Gaussian approximation does not take into account the exchange interaction\cite{Exc} (represented by the bubble diagrams), the speed of sound is overestimated by about $10\%-20\%$ compared to the value provided by the BS equation. We will also show that in the two-body problem ($\Delta=0$), the BS formalism provides the same bound-state-energy equation (\ref{2EB}) as in the Gaussian approximation. We summarize in Sec. IV.

\section {Single-particle Green's function of Rashba SOC and Zeeman field in a square optical lattice in the mean-field approximation}
   We restrict the discussion to the case of atoms confined to the lowest-energy band (single-band  model), with two possible states described by pseudospins $\sigma=\uparrow,\downarrow$. There are $M $ atoms distributed along $N$ sites, and the corresponding filling factor  $f=M/N$ is smaller than unity.  The Hamiltonian for a uniform system is $\hat{H}=\hat{H}_0+\hat{H}_{SOC}+\hat{H}_Z$, where the Hubbard Hamiltonian is
\begin{equation}\widehat{H}_0=-t\sum_{<i,j>,\sigma}\psi^\dag_{i,\sigma}\psi_{j,\sigma}
-U\sum_i \widehat{n}_{i,\uparrow}\widehat{n}_{i,\downarrow}-\mu\sum_{i,\sigma}\widehat{n}_{i,\sigma}.\label{H}
\end{equation}
Here $t=1$ is the tunneling strength of the atoms between nearest-neighbor sites, $\mu$ is the chemical potential, and $\widehat{n}_{i,\sigma}=\psi^\dag_{i,\sigma}\psi_{i,\sigma}$ is the density operator on site $i$. The Fermi operator
$\psi^\dag_{i,\sigma}$ ($\psi_{i,\sigma}$) creates (destroys) a fermion on the lattice site $i$  with pseudospin projection
$\sigma$. The symbol $\sum_{<ij>}$ means sum over nearest-neighbor sites of the 2D lattice.  The strength of the on-site interaction is $U>0$, which corresponds to attractive interaction.
 The SOC part of the Hamiltonian is given by:
\begin{equation}
\widehat{H}_{SOC}=-\imath\lambda\sum_{\langle i,j\rangle} \left( \psi^\dagger_{i,\uparrow}, \; \psi^\dagger_{j,\downarrow} \right) \left( \vec{\sigma} \times \mathbf{d}_{i,j} \right)_z
\left( {\begin{array}{*{20}c}
	{ \psi_{i,\uparrow} }  \\
	{ \psi_{j,\downarrow}}  \\
	\end{array}} \right),
\end{equation}
where $\lambda$ is the Rashba SOC coefficient, $\vec{\sigma}=(\sigma_x,\sigma_y,\sigma_z)$, the components $\sigma_{x,y,z}$ are the Pauli matrices, and $\mathbf{d}_{i,j}$ is a unit vector along the line that connects site $j$ to $i$. The out-of-plane Zeeman field is described by the term $\widehat{H}_Z$:
\begin{equation}
\widehat{H}_Z=h\sum_{i}\left( \psi^\dagger_{i,\uparrow}, \; \psi^\dagger_{i,\downarrow} \right) \sigma_z
\left( {\begin{array}{*{20}c}
	{ \psi_{i,\uparrow} }  \\
	{ \psi_{i,\downarrow}}  \\
	\end{array}} \right).
\end{equation}
\begin{figure}[htf]
	\centering\includegraphics[scale=0.33]{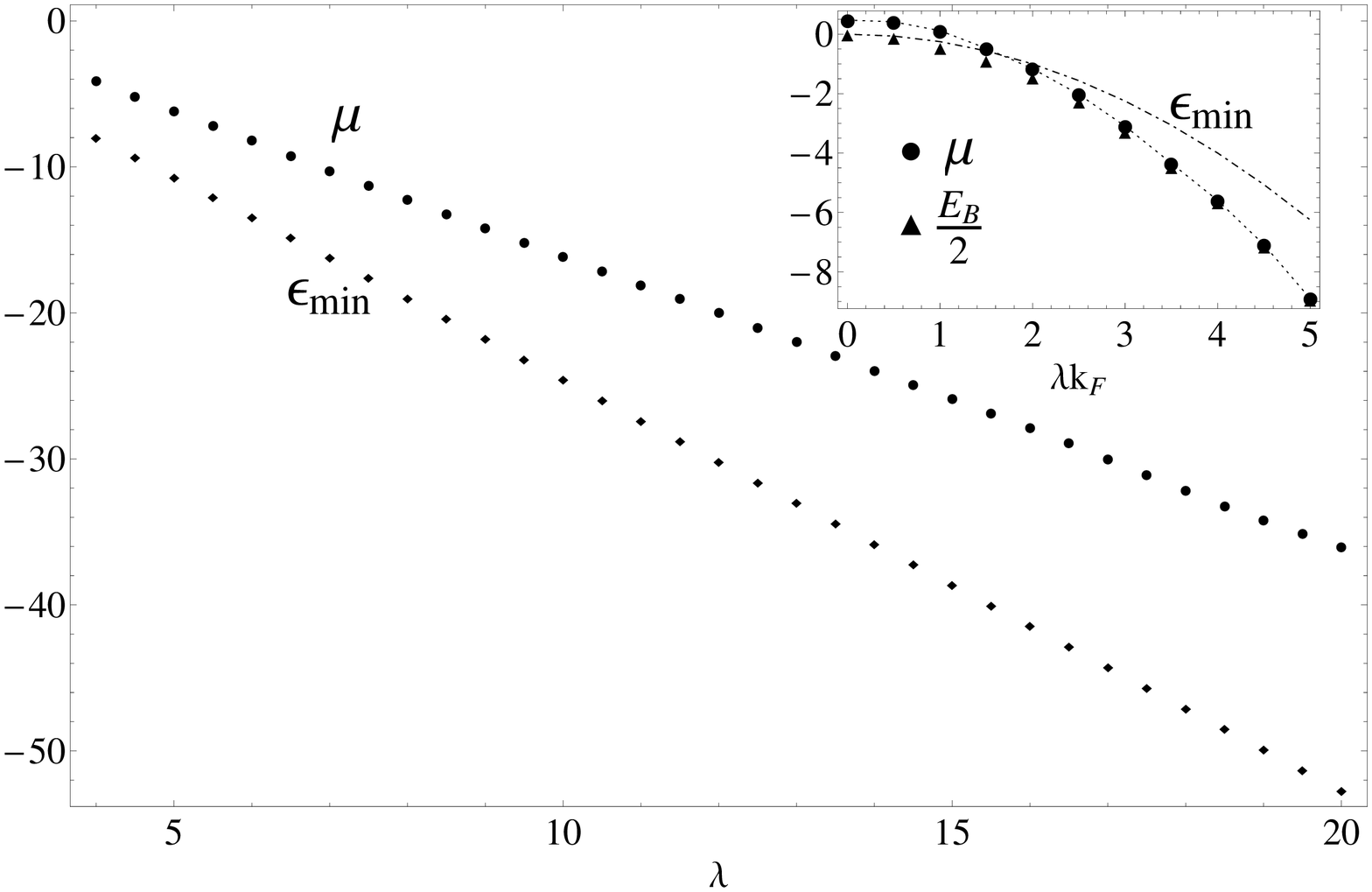}
	\caption{The  chemical potential $\mu$ and the single-particle
ground state energy $\epsilon_{min}$ as functions of the Rashba SOC strength for Fermion atoms in a 2D optical lattice. The system parameters are $U=5t$, $f=0.5$ and $h=0.4 t$ (all energies are in units of $t$). The upper insert shows the convergence of the chemical potential $\mu$ and the half of the bound state energy $E_B/2$ as functions of the Rashba SOC strength for a homogeneous unitary Fermi gas at zero temperature in the absence of a Zeeman field. The numerical values for $\mu$ and $E_B$ (in units of the Fermi energy $E_F=\hbar^2k_F^2/2m$) have been calculated in [\onlinecite{SOC6}].}\label{F4}
\end{figure}
\begin{figure}[htf]
	\centering\includegraphics[scale=0.7]{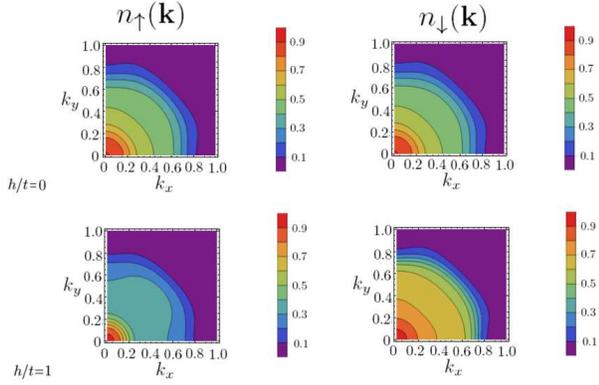}
	\caption{Zero temperature momentum distributions for the two spin components for different values of the Zeeman field $h$ at $\lambda=1t$, $f=0.5$  and $U=5t$ ($k_x$ and $k_y$ are in units of $\pi/a$). For $h=0$ we always have equal population in both spin components, but for $h>0$ the spin-up component has less population. }\label{F5}
\end{figure}
\begin{figure}[htf]
	\centering\includegraphics[scale=0.7]{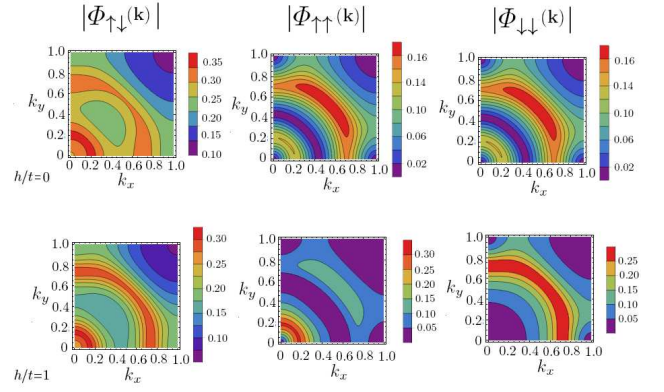}
	\caption{Magnitudes of the singlet and triplet pairing amplitudes for different values of the Zeeman field $h$ at zero temperature and at  $\lambda=1t$, $f=0.5$ and $U=5t$ ($k_x$ and $k_y$ are in units of $\pi/a$). The Zeeman field enhances the triplet pairing and reduces the singlet one.}\label{F6}
\end{figure}
\begin{figure}[htf]
	\centering\includegraphics[scale=0.7]{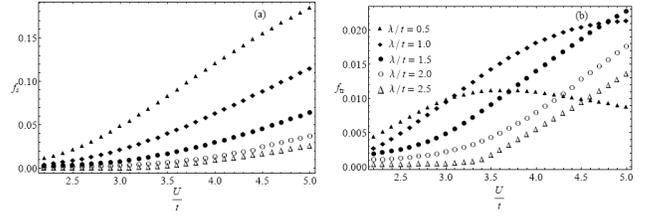}
	\caption{Zero temperature (a) singlet and (b) triplet condensate fractions for Fermion atoms in a 2D optical lattice as a function of the strength of the attractive interaction $U$ for different values of the  Rashba SOC strengths $\lambda$ at a Zeeman field $h=0.4 t$ and $f=0.5$}\label{F7}
\end{figure}
\begin{figure}[htf]
	\centering\includegraphics[scale=0.35]{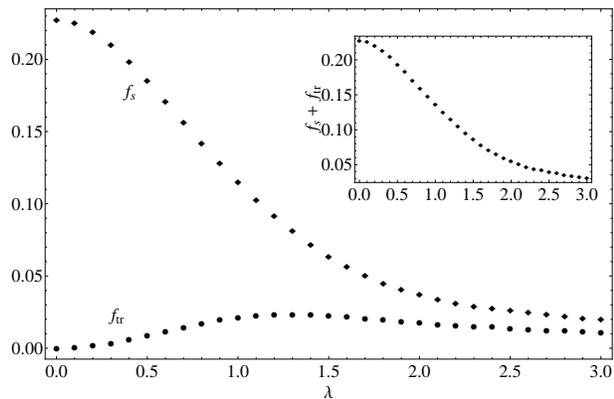}
	\caption{Zero temperature singlet $f_s$, triplet $f_{tr}$ and full $fc=f_s+f_{tr}$ condensate fractions as a function of the Rashba SOC strength $\lambda$ for Fermion atoms in a 2D optical lattice ($\lambda$ is in units of $t$, and the system parameters are the same  as in Fig. \ref{F3}). }\label{F8}
\end{figure}
\begin{figure}[htf]
	\centering\includegraphics[scale=0.25]{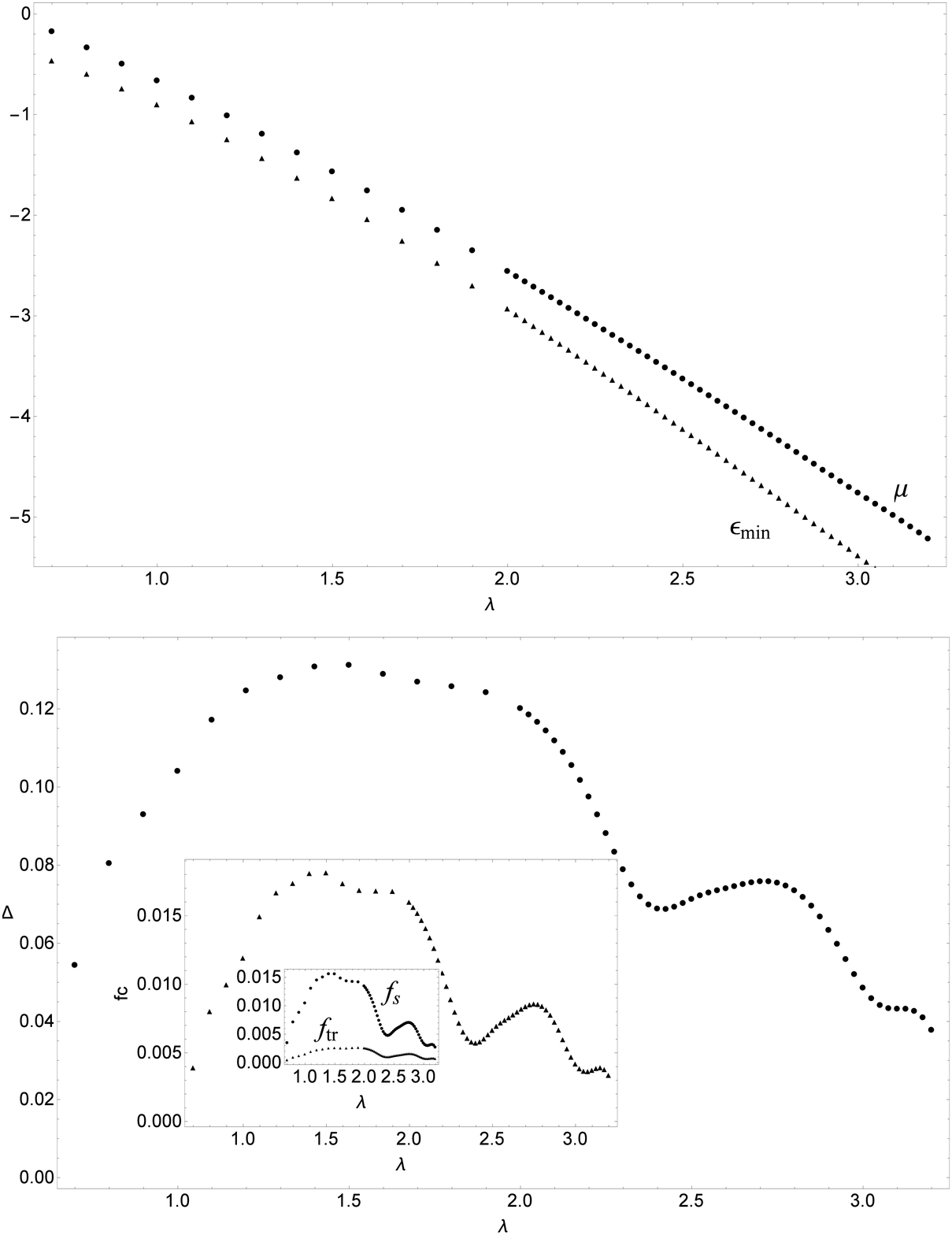}
	\caption{Zero temperature chemical potential, gap, singlet $f_s$, triplet $f_{tr}$ and full $fc=f_s+f_{tr}$ condensate fractions as a function of the Rashba SOC strength $\lambda$ for Fermion atoms in a 2D optical lattice without a Zeeman field. The system parameters are $U=2.5t$ and $f=0.1$. }\label{F9}
\end{figure}
The spectrum of the single-particle excitations in the two-body problem is $\epsilon_{\pm}(\textbf{k})=\varepsilon(\textbf{k})\pm\sqrt{h^2+4\lambda^2\left(\sin^2(k_x)+\sin^2(k_y)\right)}$, where $\varepsilon(\textbf{k})=2[1-\cos (k_x)]+2[1-\cos (k_y)]$ is the  tight-binding energy in a 2D optical lattice.

In the case when $h>h_c=2\lambda^2/t $, the lowest
single-particle state $\epsilon_{min}=-h$ is centered at $\textbf{k} = 0$.   If $h<h_c$, the lowest
single-particle state is four times degenerate, and occurs at four points in momentum space
$$\pm k_x a, \pm k_y a=\cos^{-1}\left(\frac{t\sqrt{h^2+8\lambda^2}}{\lambda\sqrt{8t^2+4\lambda^2}}\right).$$
The corresponding minimum is:
$$\epsilon_{min}=4t-\sqrt{\frac{(h^2+8\lambda^2)(2t^2+\lambda^2)}{\lambda^2}},\qquad h<h_c$$
The two-particle bound energy $E_B<0$ (in units $t$) in a 2D optical lattice  satisfies the conditions:
\begin{eqnarray}
&\frac{|E_B|}{2}>h, \quad h>h_c\nonumber \\
&\frac{|E_B|}{2}>4\sqrt{1+\frac{\lambda^2}{2}+\frac{h^2}{16}\left(1+\frac{2}{\lambda^2}\right)}-4,\quad h<h_c
\label{CEB}\end{eqnarray} When  $h<h_c$ and $\lambda>>1$, we have $|\epsilon_{min}|\propto \lambda$.

By definition, the  single-particle Green's function  $\hat{G}(x_1;y_2)$ is a $4\times 4$ matrix whose elements are $G_{\sigma,\sigma'}(x_1;y_2)=-<\widehat{T}_u\left(\psi_\sigma(x_1)\psi^\dag_{\sigma'}(y_2)\right)>$, where the symbol $<...>$ means that the thermodynamic average is
taken. Here, we have introduced composite variables,
$y_2=\{\textbf{r}_2,u_2\}$ and
$x_1=\{\textbf{r}_{1},u_1\}$, where
$\textbf{r}_{1},\textbf{r}_{2}$ are the lattice
site vectors, and according to imaginary-time (Matsubara) formalism
the variables $u_1$ and $u_2$ range from $0$ to $\hbar \beta=\hbar/(k_BT)$. Throughout
this paper we have assumed $\hbar=k_B=1$.

The single-particle properties will be studied within the standard mean-field approximation, such that the chemical potential and the pairing gap are obtained by solving the number and the gap equations self-consistently. The two equations as well as the matrix elements of the mean-field single-particle Green's function are provided in Appendix A.  The single-particle excitations are defined by the four poles $\pm \omega_1(\textbf{k})$ and $\pm \omega_2(\textbf{k})$ of the Green's function.   It is worth mentioning that Monte Carlo simulations have shown that, at zero temperature,
beyond-mean-field effects are negligible on the BCS side.\cite{MC1,MC2}

Next, we shall show various mean-field quantities of
physical interest, such as  the chemical potential, the pairing gap, the singlet and triplet pairing amplitudes, and the singlet and triplet condensate fractions. We focus on the zero-temperature case  assuming a filling factor of $f=0.5$, and a Zeeman field $h=0.4t$.  We show in Fig. \ref{F2}: (a) the chemical potential $\mu$, and (b) the gap $\Delta$ as functions of the strength of the attractive interaction $U$, for different values of the Rashba SOC strength, and will compare them to the the case of a 2D free Fermi gas \cite{SOC5} and the NAL system. \cite{OL1}

As in the case of a 2D free Fermi gas,  the chemical potential is pushed toward more negative values when the SOC increases, and  it becomes negative on the BCS side when the strength of the SOC  is increased above $\lambda_0\approx2t$. In the free case, as well as our lattice case and the NAL case, $\mu$ decreases for increasing SOC strength. However, there is little dependance of $\mu$ on interaction strength for this filling factor, $f = 0.5$.

The gap differs more significantly from the free case. At a fixed interaction, the gap is increased for higher SOC in the free case, and suppressed for higher SOC in both lattice cases. Furthermore, as the interaction goes from BCS to BEC in the free case, the gaps for the range of SOC strength's increase significantly and converge. For our lattice case, the gap decreases for the range of SOC strength's, which converge as interaction strength decreases to zero. However, our results for the gap seem to have a progression consistent with the NAL case.

In Fig. \ref{F3}, we plot the chemical potential, and the gap, as functions of the Rashba SOC strength in the weak coupling limit when $U=5t$. Contrary to the free Fermi gas,\cite{SOC5,SOC5a,SOC15} the gap decreases when SOC increases, and the chemical potential does not approach $2\epsilon_{min}$ even in the limit of very strong SOC (see Fig. \ref{F4}
). We also calculated the chemical potentials for interaction strengths U = 3t and 4t as functions of $\lambda$, but their differences to 5t values were statistically insignificant, $\mu$ still diverges from $\epsilon_{min}$, thus they have not been presented in Fig. \ref{F4}.

From the mean-field elements of the single-particle Green's function, one can obtain the momentum distribution for the two spin components
$n_{\uparrow}(\textbf{k})=<\psi^\dagger_{\textbf{k}\uparrow}\psi_{\textbf{k}\uparrow}>$, $n_{\downarrow}(\textbf{k})=<\psi^\dagger_{\textbf{k}\downarrow}\psi_{\textbf{k}\downarrow}>$. In Fig. \ref{F5}, we plot the zero temperature momentum distributions for the two spin components, for two different values of the Zeeman field. Without a Zeeman field, we have equal populations of both spin components. However, for a Zeeman field in the up-direction, the spin-up component has a smaller population consistent with the results from a two component free Fermi gas with a Zeeman field. \cite{SOC4}

Another interesting feature of the SOC, is that the pairing field contains both a singlet and
a triplet component. The  singlet $\Phi_{\downarrow\uparrow}(\textbf{k})=-\Phi_{\uparrow\downarrow}(\textbf{k})=<\psi_{\textbf{k}\downarrow}\psi_{-\textbf{k}\uparrow}>$, and  triplet  $\Phi_{\uparrow\uparrow}(\textbf{k})=<\psi_{\textbf{k}\uparrow}\psi_{-\textbf{k}\uparrow}>$, $\Phi_{\downarrow\downarrow}(\textbf{k})=<\psi_{\textbf{k}\downarrow}\psi_{-\textbf{k}\downarrow}>$  amplitudes, obtained by means of the Green's function elements $G^{MF}_{23}$, $G^{MF}_{13}$ and $G^{MF}_{24}$, are defined in Appendix A.
In Fig. \ref{F6}, we plot the magnitudes of the singlet and triplet pairing amplitudes for different values of the Zeeman field $h$ at zero temperature. As can be seen, the Zeeman field enhances the triplet pairing and reduces the singlet one consistent with free gas results with a Zeeman field. \cite{SOC4,SOC6}

With the help of the pairing amplitudes, one can calculate the condensate fraction $f_c=f_s+f_{tr}$, where $f_s$ and $f_{tr}$ are the singlet and triplet contributions, respectively. At zero temperature, we have
$f_s= \frac{2}{N}\sum_{\mathbf{k}}|\Phi_{\downarrow\uparrow}(\textbf{k})|^2$, and $f_{tr}= \frac{1}{N}\sum_{\mathbf{k}}\left(|\Phi_{\uparrow\uparrow}(\textbf{k})|^2+|\Phi_{\downarrow\downarrow}(\textbf{k})|^2\right)
$.
We show in Fig. \ref{F7} the zero temperature singlet $f_s$ and triplet $f_{tr}$  condensate fractions as functions of the strength of the attractive interaction $U$, for different values of the Rashba SOC strength. In our case and in the NAL case, the progression of the singlet condensation curves as interaction increases is consistent with the behavior of the pairing gap curves. Furthermore, the singlet condensation increases when $U$ increases; but at a fixed U, $f_s$ is suppressed by the SOC.

The  singlet $f_s$, triplet $f_{tr}$ and full $f_c$ condensate fractions versus
SOC, for interaction strength $U=5t$ are plotted in Fig. \ref{F8}. For positive chemical potential, the triplet condensation fraction is less than the singlet one. As chemical potential decreases and becomes negative ($\lambda \approx 1.96t$), the two fractions converge to each other. We see in the inset that $f_c$ decreases with respect to $\lambda$. For $\lambda=3t$, we  have $f_c$ is about $10\%$ of $f$. In the weak interaction regime, for both our case and the 2D free space without Zeeman field\cite{SOC5}, the condensate fraction is small.

Under a different scenario, such that when the Zeeman field is removed, the filling factor is reduced to $f=0.1$, and the interaction strength is weak, $U=2.5t$, the gap and the condensate fraction increase dramatically above a critical $\lambda_c$, (for our case $\lambda\approx 0.7t$), which is similar to the NAL case (see Fig. \ref{F9}). Therefore, in accordance with Ref. [\onlinecite{OL1}], formation of the SOC induced bound states takes place.

\section{Two-particle excitation spectrum in the Bethe-Salpeter approximation}
 The results in the previous Section were obtained by applying the mean-field decoupling of the quartic term in the interaction part of the Hamiltonian $\widehat{H}_0$. To go beyond the mean-field approximation, we use the idea that one can transform the quartic terms into quadratic form by making the Hubbard-Stratonovich transformation (HST) for the fermion operators. In contrast to the previous approaches, such that after performing the HST the fermion degrees of freedom are integrated out; we decouple the quartic problem by introducing a model system which consists of a multicomponent boson field $A_\alpha$ interacting with fermion fields $\psi^{\dagger}$ and $\psi$.

There are three advantages of keeping both the fermion and the boson degrees of freedom. First, the approximation that is used to decouple the self-consistent relation between the fermion self-energy and the two-particle Green's function, automatically leads to conserving approximations. This is because it relies on the fact that the BS kernel can be written as functional derivatives of the Fock $\Sigma^F$ and the Hartree $\Sigma^H$ self-energy $I=I_d+I_{exc}=\delta\Sigma^F/\delta G +\delta\Sigma^H/\delta G$. Second, the collective excitations of the Hubbard model can be calculated as poles of the fermion two-particle Green's function, $K$, and as poles of the  Green's function of the multicomponent boson field, $D$. Third, the action which describes the interactions in the Hubbard model, is similar to the action $\psi^{\dagger}A\psi$ in quantum electrodynamics. For this reason we can utilize powerful field-theoretical methods, such as the method of Legendre transforms, to derive the Schwinger-Dyson (SD) equation of the boson Green's function $\widehat{D}$, the BS equation of the two-particle Green's function, and the corresponding equation for the vertex function.

Thus, in the aforementioned manner, the quartic terms are  transformed into the quadratic form by inserting a four-component boson field $A_\alpha(z)$ which mediates the interaction of fermions $\widehat{\overline{\psi}} (y)=\widehat{\Psi}^\dag (y)/\sqrt{2}$ and $\widehat{\psi}(x)=\widehat{\Psi}(x)/\sqrt{2}$ $(\alpha=1,2,3,4$, $z=(\textbf{r}_i,v)$, where

$$\widehat{\Psi}(x)=\left(
\begin{array}{c}
\psi_\uparrow(x) \\
\psi_\downarrow(x) \\
\psi^\dag_\uparrow(x)\\
\psi^\dag_\downarrow(x)\\
\end{array}%
\right),$$ $$ \widehat{\Psi}^\dag
(y)=\left(\psi^\dag_\uparrow(y),\;\psi^\dag_\downarrow(y),\;
\psi_\uparrow(y),\;\psi_\downarrow(y)
\right).
$$

The single-particle Green's function, introduced in the previous section, can be written as a  tensor product between the two matrices $\widehat{\Psi}(x_1)$ and $\widehat{\Psi}^\dag
(y_2)$, i.e. $\widehat{G}(x_1;y_2)=-\langle\widehat{T}_u(\widehat{\Psi}(x_1)\otimes\widehat{\overline{\Psi}}(y_2))\rangle$.  As in quantum electrodynamics, where the photons mediate the interaction of electric charges, we define an action of the following form $S= S^{(F)}_0+S^{(B)}_0+S^{(F-B)}$, where $S^{(F)}_0=\widehat{\overline{\psi}}(y)\widehat{G}^{(0)-1}(y;x)\widehat{\psi} (x)$, $S^{(B)}_0=\frac{1}{2}A_{\alpha}(z)D^{(0)-1}_{\alpha \beta}(z,z')A_{\beta}(z')$, and $S^{(F-B)}=\widehat{\overline{\psi}}
(y)\widehat{\Gamma}^{(0)}_{\alpha}(y,x\mid z)\widehat{\psi}(x)A_{\alpha}(z).$ The action $S_0^{(F)}$ describes the fermion part of the system. The generalized inverse Green's function of free fermions $\widehat{G}^{(0)-1}(y;x)$ is given by the following $4\times4$ matrix:

\begin{displaymath}\widehat{G}^{(0)-1}(y;x)=\sum_{\mathbf{k},\omega_m}e^{\left[\imath
\mathbf{k}\cdot(\mathbf{r}_i-\mathbf{r}_{i'})-\omega_m(u-u')\right]}
\widehat{G}^{(0)-1}(\mathbf{k},\imath\omega_m).
\end{displaymath}
 In the case of the population-balanced Fermi gas with a planar Rashba SOC and an out-of-plane Zeeman field, the noninteracting Green's function is

\begin{displaymath}
\widehat{G}^{(0)-1}(\mathbf{k},\imath\omega_m)= \left(
\begin{array}{cc}
g^{(0)-1}_{11} & 0  \\
0 & g^{(0)-1}_{22} \\
\end{array}
\right),
\end{displaymath} \label{zTspGFMF}
where
\begin{widetext}
\begin{displaymath}
g^{(0)-1}_{11}=\left(
\begin{array}{cc}
\imath\omega_m-\xi(\mathbf{k})-h & -2\lambda(\sin k_y + \imath\sin k_x) \\
-2\lambda(\sin k_y - \imath\sin k_x) & \imath\omega_m-\xi(\mathbf{k}) + h \\
\end{array}
\right) \;
\end{displaymath}
\begin{displaymath}
g^{(0)-1}_{22}=\left(
\begin{array}{cc}
\imath\omega_m + \xi(\mathbf{k}) + h & -2\lambda(\sin k_y - \imath\sin k_x) \\
-2\lambda(\sin k_y + \imath\sin k_x) & \imath\omega_m + \xi(\mathbf{k}) - h \\
\end{array}
\right).
\end{displaymath}
\end{widetext}
The action $S^{(B)}_0$ describes the boson field which mediates the fermion-fermion on-site interaction in the Hubbard Hamiltonian. The bare boson propagator $\widehat{D}^{(0)}$ is defined as

\begin{displaymath}
\widehat{D}^{(0)}
(z,z')=\delta(v-v')U\delta{j,j'}\left(%
\begin{array}{cccc}
0&1&0&0  \\
1 &0 &0&0  \\
0 &0 &0&0\\
0 &0 & 0&0
\end{array}
\right).
\end{displaymath}

The Fourier transform of the above boson propagator is given by

\begin{displaymath}
\widehat{D}^{(0)} (z,z')=\frac{1}{N} \sum_\mathbf{k}\sum_{\omega_p} e^{\left\{\imath\left[\mathbf{k}\cdot\left(\mathbf{r}_j-\mathbf{r}_{j'}\right)	 -\omega_p\left(v-v'\right)\right]\right\}}\widehat{D}^{(0)}(\mathbf{k}),
\end{displaymath}

\begin{displaymath}
\widehat{D}^{(0)}(\mathbf{k})= \left(
\begin{array}{cccc}
0&U &0&0  \\
U  &0 &0&0  \\
0 &0 &0&0 \\
0 &0 & 0&0
\end{array}%
\right).
\end{displaymath}
Here the symbol $\sum_{\omega_p}$ is used to denote $\beta^{-1}\sum_{p}$ (for a boson field $\omega_p=
   (2\pi/\beta)p ;p=0, \pm 1, \pm 2,... $).

The interaction between the fermion and the boson fields is described by the action $S^{(F-B)}$. The bare vertex
$\widehat{\Gamma}^{(0)}_{\alpha}(y_1;x_2\mid z)=\widehat{\Gamma}^{(0)}_{\alpha}(i_1,u_1;i_2, u_2\mid
j,v)=\delta(u_1-u_2)\delta(u_1-v)\delta_{i_1i_2}\delta_{i_1j}\widehat{\Gamma}^{(0)}(\alpha)$ is a $4\times 4$ matrix, where

\begin{equation}\begin{split}
&\widehat{\Gamma}^{(0)}(\alpha)=\frac{1}{2}(\gamma_0+\alpha_z)\delta_{\alpha1}
+\frac{1}{2}(\gamma_0-\alpha_z)\delta_{\alpha2}+\\
&\frac{1}{2}(\alpha_x+\imath\alpha_y)\delta_{\alpha3}+
\frac{1}{2}(\alpha_x-\imath\alpha_y)\delta_{\alpha4}.
\end{split}\end{equation} \label{Gamma0}

The Dirac matrix $\gamma_0$,  and the  matrices $\widehat{\alpha}_i$, are defined as follows; such that when a four-dimensional space is used, the electron spin operators $\sigma_i$ must be replaced by $\widehat{\alpha}_i  \gamma_0$:\cite{M}

\begin{displaymath}
\gamma_0=\left(
\begin{array}{cccc}
1&0&0&0  \\
0&1&0&0  \\
0& 0& -1&0  \\
0& 0& 0&-1  \\
\end{array}%
\right),  \widehat{\alpha}_i=\left(%
\begin{array}{cc}
\sigma_i & 0  \\
0& \sigma_y\sigma_i\sigma_y \\
\end{array}%
\right), i=x,y,z.
\end{displaymath}

The relationship between the Hubbard model, and our model system, can be demonstrated by applying the Hubbard-Stratonovich transformation (HST) for the fermion operators:
\begin{widetext}
\begin{equation}\int D\mu[A]\exp\left[\widehat{\overline{\psi}}
(y)\widehat{\Gamma}^{(0)}_{\alpha}(y;x|z)\widehat{\psi}(x)A_{\alpha}(z)\right]
=\exp\left[-\frac{1}{2}\widehat{\overline{\psi}}
(y)\widehat{\Gamma}^{(0)}_{\alpha}(y;x|z)\widehat{\psi}(x)
D_{\alpha,\beta}^{(0)}(z,z') \widehat{\overline{\psi}}
(y')\widehat{\Gamma}^{(0)}_{\beta}(y';x'|z')\widehat{\psi}(x')\right].\label{HSa}
\end{equation}
\end{widetext}
The functional measure $D\mu[A]$ is chosen to be:

\begin{displaymath}
D\mu[A]=DAe^{-\frac{1}{2}A_{\alpha}(z)D_{\alpha,\beta}^{(0)-1}(z,z')
	A_{\beta}(z')},\int D\mu[A] =1.
\end{displaymath}

Next, we shall apply a functional integral technique to derive the BS equation for the poles (collective modes $\omega=\omega(\textbf{Q})$) of the two-particle Green's function in the GRPA.  By following the standard procedures in the functional integral formalism \cite{BS1,BS2}, it is possible to derive the SD  equation $G^{-1}=G^{(0)-1}-\Sigma$, and the BS  equation $[K^{(0)-1}-I]\Psi=0$ for the poles of the single-particle Green's function $G$, and the poles of the two-particle Green's function, respectively.\cite{FI1,FI2,BetheS} Here, $G^{(0)}$ is the free single-particle propagator, $\Sigma$ is the fermion self-energy, $I$ is the BS kernel, and the two-particle free propagator $K^{(0)}=GG$ is a product of two fully dressed single-particle Green's functions. The kernel of the BS equation is defined as a sum of the direct interaction $I_d=\delta\Sigma^F/\delta G$, and the exchange interaction $I_{exc}=\delta\Sigma^H/\delta G$, where $\Sigma^F$ and $\Sigma^H$ are the Fock and the Hartree parts of the fermion self-energy $\Sigma$. Since the fermion self-energy depends on the two-particle Green's function, the positions of both poles must be obtained by resolving the SD and BS equations self-consistently.

In practice, the GRPA permits us to decouple the SD equations from the BS equation. According to this approximation, the single-particle excitations are calculated in the mean field approximation, while the collective modes are obtained by solving the BS equation in which the single-particle Green's functions are determined in Hartree-Fock approximation, and the BS kernel takes into account contributions from both ladder diagrams (direct interaction) and bubble diagrams (exchange interaction). Explicitly, the single-particle mean-field Green's function is

\begin{displaymath}
\widehat{G}^{-1}_{MF}(\mathbf{k},\imath\omega_m)= \left(
\begin{array}{cc}
g^{(0)-1}_{11} & \imath \Delta\sigma_y  \\
-\imath \Delta\sigma_y & g^{(0)-1}_{22} \\
\end{array}
\right).
\end{displaymath} \nonumber

In the GRPA the direct interaction in the BS kernel is calculated by a linearized Fock term and exact Hartree term:
\begin{widetext}
\begin{equation}
\Sigma^F_0(i_1,u_1;i_2,u_2)_{n_1,n_2}=\nonumber \\
-U\delta_{i_1,i_2}\delta(u_1-u_2)
\left(
\begin{array}{cccc}
0 & G_{12} & 0 & -G_{14} \\
G_{21} & 0 & -G_{23} & 0 \\
0 & G_{32} & 0 & G_{34} \\
-G_{41} & 0 & G_{43} & 0 \\
\end{array}
\right),
\end{equation}

\begin{equation}
\Sigma^H(i_1,u_1;i_2,u_2)= \frac{U}{2}\delta_{i_1,i_2}\delta(u_1-u_2)
\left(
\begin{array}{cccc}
G_{22}-G_{44} & 0 & 0 & 0 \\
0 & G_{11}-G_{33} & 0 & 0 \\
0 & 0 & G_{44}-G_{22} & 0 \\
0 & 0 & 0 & G_{33}-G_{11} \\
\end{array}
\right),
\end{equation}
where $G_{i,j}\equiv G_{i,j}(1;2)=G_{i,j}(i_1,u_1;i_2,u_2)$. In the GRPA the BS equation for the sixteen BS amplitudes $\Psi^{\mathbf{Q}}_{n_2,n_1}$, $\{n_1,n_2\}=\{1,2,3,4\}$, is
\begin{displaymath}\Psi^{\mathbf{Q}}_{n_2n_1}=K^{(0)}\left(
\begin{array}{cc}
n_1 & n_3  \\
n_2 & n_4 \\
\end{array}
|\omega(\mathbf{Q})\right)
\left[I_d\left(
\begin{array}{cc}
n_3 & n_5  \\
n_4 & n_6 \\
\end{array}%
\right)+I_{exc}\left(
\begin{array}{cc}
n_3 & n_5  \\
n_4 & n_6 \\
\end{array}
\right)\right]\Psi^{\mathbf{Q}}_{n_6,n_5},
\end{displaymath} \label{BSEdZ1}
\end{widetext}
where
\begin{displaymath}
I_d\left(
\begin{array}{cc}
n_1 & n_3  \\
n_2 & n_4 \\
\end{array} \right) =-\Gamma^{(0)}_\alpha(n_1,n_3)D^{(0)}_{\alpha\beta}\Gamma^{(0)}_\beta(n_4,n_2),
\end{displaymath}
\begin{displaymath}
I_{exc}\left(
\begin{array}{cc}
n_1 & n_3  \\
n_2 & n_4 \\
\end{array} \right) = \frac{1}{2}\Gamma^{(0)}_\alpha(n_1,n_2)D^{(0)}_{\alpha\beta}\Gamma^{(0)}_\beta(n_4,n_3),
\end{displaymath}
are the direct and exchange interactions, correspondingly. The two-particle free propagator $K^{(0)}$ in the GRPA is defined as follows:

\begin{equation}\begin{split}&K^{(0)}\left(
\begin{array}{cc}
n_1 & n_3  \\
n_2 & n_4 \\
\end{array}
|\omega(\mathbf{Q})\right)\equiv Kn_1n_3n_4n_2=\nonumber\\&\int
\frac{d\Omega}{2\pi}
\int\frac{d^d\mathbf{k}}{(2\pi)^d}G^{MF}_{n_1n_3}
\left(\mathbf{k}+\mathbf{Q},\Omega+\omega(\mathbf{Q})\right)G^{MF}_{n_4n_2}(\mathbf{k},\Omega).
\end{split}\end{equation}

The above BS equation can be rewritten in matrix form as $(\widehat{I}+U\widehat{M}(\omega,\textbf{Q}))\widehat{\Psi}=0$, where $\widehat{I}$ is the unit matrix, and the transposed matrix of $\widehat{\Psi}$ is given by

\begin{eqnarray}
\widehat{\Psi}^T=\left(\Psi^{\mathbf{Q}}_{1,1}, \; \Psi^{\mathbf{Q}}_{1,2}, \; \Psi^{\mathbf{Q}}_{1,3}, \; \Psi^{\mathbf{Q}}_{1,4}, \; \Psi^{\mathbf{Q}}_{2,1}, \; \Psi^{\mathbf{Q}}_{2,2}, \; \Psi^{\mathbf{Q}}_{2,3}, \; \Psi^{\mathbf{Q}}_{2,4}, \right.  \nonumber \\
\left.
\Psi^{\mathbf{Q}}_{3,1}, \; \Psi^{\mathbf{Q}}_{3,2}, \; \Psi^{\mathbf{Q}}_{3,3}, \; \Psi^{\mathbf{Q}}_{3,4}, \; \Psi^{\mathbf{Q}}_{4,1}, \; \Psi^{\mathbf{Q}}_{4,2}, \; \Psi^{\mathbf{Q}}_{4,3}, \; \Psi^{\mathbf{Q}}_{4,4} \right). \nonumber
\end{eqnarray}
The non-trivial solution of the BS equation requires that the $16\times16$ determinant  $det|\widehat{I}+U\widehat{M}(\omega,\textbf{Q}))|=0$, and this condition provides the collective excitation spectrum $\omega(\textbf{Q})$. By applying simple matrix algebra, the $16\times16$ determinant $det|\widehat{I}+U\widehat{M}(\omega,\textbf{Q}))|=0$ can be simplified to a $10\times10$ determinant. Thus, the collective excitation spectrum $\omega(\textbf{Q})$ at a given momentum $\textbf{Q}$ is defined by $Z(\omega,\textbf{Q})=0$, where
\begin{equation}
Z(\omega,\mathbf{Q})= det\left|
\begin{array}{cc}
G_{2\times2}(\omega,\mathbf{Q}) &  D_{2\times 8}(\omega,\mathbf{Q}) \\
D_{8\times2}(\omega,\mathbf{Q}) & C_{8\times8}(\omega,\mathbf{Q})\\
\end{array}
\right|.\label{Z1010}
\end{equation}
Here  $D_{2\times 8}(\omega,\mathbf{Q})=\left[D_{8\times2}(\omega,\mathbf{Q})\right]^T$, and the elements of the  blocks $G_{2\times2}(\omega,\mathbf{Q})$, $B_{8\times 2}(\omega,\mathbf{Q})$ and
$C_{8\times8}(\omega,\mathbf{Q})$ are defined in Appendix B.

In the Gaussian approximation, the collective excitation spectrum $\omega_G(\textbf{Q})$ is defined by the equation $G_G(\omega,\mathbf{Q})=0$, where:
\begin{equation}
G_G(\omega,\mathbf{Q})=det|G_{2\times2}(\omega,\mathbf{Q})|.\label{G}\end{equation}
\begin{figure}[htf]
	\centering\includegraphics[scale=0.4]{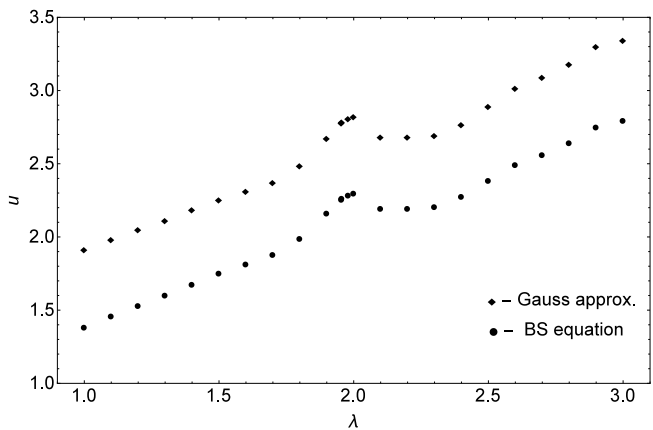}
	\caption{The speeds of the sound $u$ along the $Q_x$ direction calculated in the Gaussian (circles)   and in the Bethe-Salpeter (squares) approximations as  functions of the Rashba SOC strength $\lambda$. The SOC strength is in units of $t$, while the speed of sound is in units of $ta/\hbar$. The system parameters are the same as in Fig. \ref{F3}. }\label{F10}\end{figure}
In the BS approach, when  all single particle Green's functions are taken into account, the collective excitation spectrum $\omega_{BS}(\textbf{Q})$ is defined by the equation $G_{BS}(\omega,\mathbf{Q})=0$, where
\begin{equation}\begin{split}
&G_{BS}(\omega,\mathbf{Q})=\\&det|G_{2\times2}(\omega,\mathbf{Q})-D_{2\times8}(\omega,\mathbf{Q})
C^{-1}_{8\times8}(\omega,\mathbf{Q})D_{8\times2}(\omega,\mathbf{Q})|.\label{BS}\end{split}\end{equation}
In Fig. \ref{F10}, we have plotted the speed of sound along x-direction ($\textbf{Q} = (Q_x , 0)$) as a
function of the SOC strength, calculated within the Gaussian approximation and from
the BS equation.  Close to the point where the chemical potential becomes negative, the speed of sound has a local maximum. As can be seen, the Gaussian approximation overestimates the speed of sound, and the difference between  the two approaches is about $15\%$. This can be related to the fact that the Gaussian approximation does not take into account the exchange interaction. The speed of sound  exhibits different behavior compared to the continuum: in the 3D and 2D continuum the slope of the Goldstone mode decreases as a function of the SOC strength,\cite{SOC11} while in the lattice case the speed of sound increases monotonically with SOC strength.

Next, we shall apply the BS formalism to derive an equation for the  bound state energy, and show that it is equivalent to the bound state equation in the Gaussian approximation when $\Delta = 0$. In both cases, we set $\textbf{Q}=0$,  and the two-body problem follows from our many-body description by setting $\omega-\mu=E_B$. In this limit, the Gaussian block $G_{2\times 2}$ assumes a diagonal form:
\begin{equation}
G_{2\times2}(\omega,\mathbf{Q})=\left|
\begin{array}{cc}
g_{G}(\omega,\mathbf{Q}) &  0 \\
0 & g_{G}(-\omega,\mathbf{Q})\\
\end{array}
\right|, \label{2G}\end{equation}
where  \begin{equation}\begin{split}&g_{G}(\omega,\mathbf{Q})\\&=1+\frac{U}{2}\left[K_{1144}(\omega,\mathbf{Q})+K_{2233}(\omega,\mathbf{Q})-2K_{1234}(\omega,\mathbf{Q})\right] .\nonumber\end{split}\end{equation}
\begin{figure}[htf]
	\centering\includegraphics[scale=0.36]{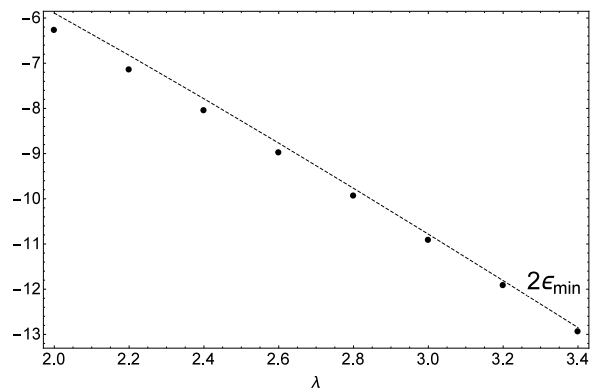}
	\caption{The bound-state energy $E_B$ as  functions of the Rashba SOC strength $\lambda$ at $U=5t$ and  $h=0.4t$. The dashed line is twice the single-particle minimum energy $2\epsilon_{min}$ ( the SOC strength and all energies are in units of $t$).}\label{F11}\end{figure}
 Thus, in the Gaussian approximation the determinant   $det|G_{2\times2}(\omega=E_B+\mu,\mathbf{Q}=0)|=0$ provides the bound-state equation at zero temperature (\ref{2EB}). The BS equation for the bound-state energy follows from Eq. (\ref{BS}); we can obtain the following BS equation for the bound-state energy $g_{BS}(E_B+\mu,0)=0$, where
 \begin{eqnarray}&g_{BS}(\omega,\mathbf{Q})= \left(1+U\left[K_{1144}(\omega,\mathbf{Q})-K_{2233}(\omega,\mathbf{Q})\right]\right)\times\nonumber\\
&\left(1+U\left[K_{1144}(\omega,\mathbf{Q})+K_{2233}(\omega,\mathbf{Q})\right]\right)
  .\label{2BS}\end{eqnarray}

As it can be seen, if $\Delta=0$ we have $K_{1144}(\omega,\textbf{Q})=K_{2233}(\omega,\textbf{Q})$, and therefore, we re-obtain the two-body bound-state equation (\ref{2EB}) in the Gaussian approximation.

We have numerically solved the bound-state equation (\ref{2EB}) for different strengths of SOC, assuming that the  Zeeman field is $h=0.4t$ and $U=5t$, and plotted the results in Fig. \ref{F11}. As in the continuum, an increase of the SOC strength $\lambda$
leads to a deeper bound state which approaches $2\epsilon_{min}$.
\section{Summary}
In summary, we have presented a comprehensive study of the single-particle and collective excitations of
  a Rashba spin-orbit coupled atomic Fermi gas with attractive interaction, loaded in a 2D square optical lattice, in the presence of an effective out-of-plane  Zeeman field. We calculated
important and experimentally relevant physical quantities, such as the chemical potential, the pairing gap,  the singlet and triplet pairing correlations,  singlet and triplet condensate fractions, and the speed of sound. All of these quantities are strongly modified compared to  the Fermi gases in the free space. In particular, we have found entirely different behavior (compared to the continuum) of the gap,  the condensate fraction, and the speed of sound  as functions of the SOC strength. Whereas mean field quantities are similar to that of the non-Abelian SOC lattice case, even though that case did not have a Zeeman field. To the best of our knowledge, there is no other calculation of the speed of sound in a lattice case as a function of SOC strength that we can compare. Lastly, we have shown that our Bethe-Salpeter formalism is equivalent to the Gaussian approximation for $\Delta = 0$.

\appendix
\section{Mean-field single-particle Green's function}
 Under the Nambu spinor basis $\widehat{\Psi}=(\psi_{\mathbf{k},\uparrow},\,\psi_{\mathbf{k},\downarrow},\,\psi^\dagger_{-\mathbf{k},\uparrow},\,\psi^\dagger_{-\mathbf{k},\downarrow})^T$ for the field operators, the Fourier transform of the single-particle mean-field Green's function $\hat{G}^{MF}(x_1,y_2)$ is as follows:
$$
\widehat{G}^{MF}(\textbf{k},\imath \omega_m)=\left( {\begin{array}{*{20}c}
	\hat{g}(\textbf{k},\imath \omega_m)	& \hat{f}(\textbf{k},\imath \omega_m) \\
	\left(\hat{f}(\textbf{k},-\imath \omega_m)\right)^\dag & -\left(\hat{g}(-\textbf{k},-\imath \omega_m)\right)^T\\
	\end{array}} \right),
$$
where for fermion fields $\omega_m=
   (2\pi/\beta)(m +1/2) ;m=0, \pm 1, \pm 2,... $, and
$$\hat{g}(\textbf{k},\imath \omega_m)=\left( {\begin{array}{*{20}c}
	G^{MF}_{11}(\textbf{k},\imath \omega_m)	& G^{MF}_{12}(\textbf{k},\imath \omega_m) \\
	G^{MF}_{21}(\textbf{k},\imath \omega_m) &G^{MF}_{22}(\textbf{k},\imath \omega_m)\\
	\end{array}} \right),$$
$$\hat{f}(\textbf{k},\imath \omega_m)=\left( {\begin{array}{*{20}c}
	G^{MF}_{13}(\textbf{k},\imath \omega_m)	& G^{MF}_{14}(\textbf{k},\imath \omega_m) \\
	G^{MF}_{23}(\textbf{k},\imath \omega_m) &G^{MF}_{24}(\textbf{k},\imath \omega_m)\\
	\end{array}} \right).$$ \begin{widetext}
The matrix elements of $\hat{g}(\textbf{k},\imath \omega_m)$ and $\hat{f}(\textbf{k},\imath \omega_m)$ are given by
$$G^{MF}_{ij}(\textbf{k},\imath \omega_m)=\frac{A_{ij}(\textbf{k},\omega_1,\omega_2)}{\imath \omega_m-\omega_1(\textbf{k})}+\frac{B_{ij}(\textbf{k},\omega_1,\omega_2)}{\imath \omega_m+\omega_1(\textbf{k})}+\frac{A_{ij}(\textbf{k},\omega_2,\omega_1)}{\imath \omega_m-\omega_2(\textbf{k})}+\frac{B_{ij}(\textbf{k},\omega_2,\omega_1)}{\imath \omega_m+\omega_2(\textbf{k})}.$$
The the mean-field single-particle excitations are defined by the four poles $\pm \omega_1(\textbf{k})$ and $\pm \omega_2(\textbf{k})$, where
$$\omega_{1}(\textbf{k})=\left[S(\textbf{k})+\Delta^2+h^2+\xi^2(\textbf{k}) - 2\sqrt{h^2\Delta^2+
\xi^2(\textbf{k})\left[h^2+S(\textbf{k})\right]}\right]^{1/2},$$ $$\omega_{2}(\textbf{k})=\left[S(\textbf{k})+\Delta^2+h^2+\xi^2(\textbf{k}) +2 \sqrt{h^2\Delta^2+
\xi^2(\textbf{k})\left[h^2+S(\textbf{k})\right]}\right]^{1/2}$$
 \end{widetext}
Here, $\Delta$ is the mean-field gap, $\xi(\textbf{k})=2(1-\cos k_x)+2(1-\cos k_y)-\mu$ is the tight-binding energy,  $S(\textbf{k})= |J(\textbf{k})|^2$, such that $J(\textbf{k})=2\lambda\left[\sin(k_y)+\imath\sin(k_x)\right]$ is the Rashba interaction. The explicit expressions of the elements of the $\hat{g}(\textbf{k},\imath \omega_m)$ and $\hat{f}(\textbf{k},\imath \omega_m)$ are as follows (the $\omega$-dependence of $M(\omega_1,\omega_2)$, $A_{ij}(\omega_1,\omega_2)$ and $B_{ij}(\omega_1,\omega_2)$, as well as the $\textbf{k}$-dependence of $S(\textbf{k})$,  $J(\textbf{k})$, $\xi(\textbf{k})$, and $\omega_{1,2}(\textbf{k})$ both are understood):
\begin{widetext}
$$ M=2 (\omega^2_1-\omega^2_2)\omega_1;$$
$$M A_{11}=-(h-\xi)(h^2-\Delta^2+S-\xi^2)-\left[\Delta^2+S+(h-\xi)^2\right]\omega_1+(h+\xi)\omega_1^2+\omega_1^3;$$
$$M B_{11}=(h-\xi)(h^2-\Delta^2+S-\xi^2)-\left[\Delta^2+S+(h-\xi)^2\right]\omega_1-(h+\xi)\omega_1^2+\omega_1^3;$$
$$M A_{22}=(h+\xi)(h^2-\Delta^2+S-\xi^2)-\left[\Delta^2+S+(h+\xi)^2\right]\omega_1+(-h+\xi)\omega_1^2+\omega_1^3;$$
$$M B_{22}=-(h+\xi)(h^2-\Delta^2+S-\xi^2)-\left[\Delta^2+S+(h+\xi)^2\right]\omega_1+(h-\xi)\omega_1^2+\omega_1^3;$$
$$M A_{12}=-J\left[h^2+S+(\Delta-\xi-\omega_1)(\Delta+\xi+\omega_1)\right];$$
$$M B_{12}=J\left[h^2+S+(\Delta+\xi-\omega_1)(\Delta-\xi-\omega_1)\right];M A_{21}=-J^*\left[h^2+S+(\Delta-\xi-\omega_1)(\Delta+\xi+\omega_1)\right];$$
$$M B_{21}=J^*\left[h^2+S+(\Delta+\xi-\omega_1)(\Delta-\xi+\omega_1)\right];M A_{13}=2J\Delta(\xi-h);\quad MB_{13}=2J\Delta(h-\xi);$$
$$M A_{24}=-2J^*\Delta(\xi+h);\quad M B_{24}=2J^*\Delta(\xi+h);M A_{14}=\Delta\left[S+\Delta^2+\xi^2-(h+\omega_1)^2\right];$$
$$M B_{14}=-\Delta\left[S+\Delta^2+\xi^2-(h-\omega_1)^2\right];M A_{23}=-\Delta\left[S+\Delta^2+\xi^2-(h-\omega_1)^2\right];$$
$$M B_{23}=\Delta\left[S+\Delta^2+\xi^2-(h+\omega_1)^2\right];$$

In the system under consideration, the number of particles is fixed, so the suitable thermodynamic potential is the Helmholtz free energy (HFE) given by
$$
F(\Delta,f)=\frac{\Delta^2}{U} + \frac{1}{N}\sum_{\mathbf{k}} \left[ \xi(\mathbf{k})-\frac{1}{2}\left( \omega_1(\textbf{k}) + \omega_2(\textbf{k}) \right)   -\frac{1}{\beta} \left\{ \ln \left(1+e^{-\beta\omega_1(\textbf{k})}\right) + \ln\left(1+e^{-\beta\omega_2(\textbf{k})}\right) \right\} \right] + f\mu.$$
By minimizing the HFE with respect to the chemical potentials $\mu$ and the gap $\Delta$, we obtain a set of two equations, namely, the number and gap equations:
\begin{equation}\begin{split}
&f=1-\frac{1}{N}\sum_{\mathbf{k}} \left[ \frac{1}{2}-f_F(\omega_1(\mathbf{k})) \right] \frac{\xi(\mathbf{k})}{\omega_1(\mathbf{k})} \left( 1+\frac{S(\mathbf{k})+h^2}{\sqrt{h^2\Delta^2+
\xi^2(\textbf{k})\left[h^2+S(\textbf{k})\right]}} \right)  \nonumber \\
&-\frac{1}{N}\sum_{\mathbf{k}}\left[ \frac{1}{2}-f_F(\omega_2(\mathbf{k})) \right]\frac{\xi(\mathbf{k})}{\omega_2(\mathbf{k})} \left( 1 - \frac{S(\mathbf{k})+h^2}{\sqrt{h^2\Delta^2+
\xi^2(\textbf{k})\left[h^2+S(\textbf{k})\right]}} \right),
\\
&1=\frac{U}{N}\sum_{\mathbf{k}} \left[ \frac{1}{2}-f_F(\omega_1(\mathbf{k})) \right] \frac{1}{2\omega_1(\mathbf{k})} \left( 1+\frac{h^2}{\sqrt{h^2\Delta^2+
\xi^2(\textbf{k})\left[h^2+S(\textbf{k})\right]}} \right) + \nonumber \\
&\frac{1}{N}\sum_{\mathbf{k}}\left[ \frac{1}{2}-f_F(\omega_2(\mathbf{k})) \right]\frac{1}{2\omega_2(\mathbf{k})} \left( 1 - \frac{h^2}{\sqrt{h^2\Delta^2+
\xi^2(\textbf{k})\left[h^2+S(\textbf{k})\right]}} \right).
\end{split}\end{equation}
Here $f_F(z)=[\exp(\beta z)+1]^{-1}$ is the Fermi distribution function.
Using the corresponding matrix elements of the mean-field single-particle Green's function, we can evaluate the momentum distribution for the two spin components
$n_{\uparrow}(\textbf{k})=<\psi^\dagger_{\textbf{k}\uparrow}\psi_{\textbf{k}\uparrow}>$, $n_{\downarrow}(\textbf{k})=<\psi^\dagger_{\textbf{k}\downarrow}\psi_{\textbf{k}\downarrow}>$:
\begin{equation}\begin{split}
&n_\uparrow(\textbf{k})=1/2-\left[1/2-f_F(\omega_1(\textbf{k})\right]\frac{\xi(\textbf{k})}{2\omega_1(\textbf{k})}
\left[1-\frac{h^2+S(\textbf{k})}{\sqrt{S(\textbf{k})\xi^2(\textbf{k})+h^2\left[\Delta^2+\xi^2(\textbf{k})\right]}}\right]\nonumber\\
&-\left[1/2-f_F(\omega_2(\textbf{k})\right]\frac{\xi(\textbf{k})}{2\omega_2(\textbf{k})}
\left[1+\frac{h^2+S(\textbf{k})}{\sqrt{S(\textbf{k})\xi^2(\textbf{k})+h^2\left[\Delta^2+\xi^2(\textbf{k})\right]}}\right]\nonumber\\&-\left[1/2-f_F(\omega_1(\textbf{k})\right]\frac{h}{2\omega_1(\textbf{k})}
\left[1-\frac{\Delta^2+\xi^2(\textbf{k})}{\sqrt{S(\textbf{k})\xi^2(\textbf{k})+h^2\left[\Delta^2+\xi^2(\textbf{k})\right]}}\right]\nonumber\\
&-\left[1/2-f_F(\omega_2(\textbf{k})\right]\frac{h}{2\omega_2(\textbf{k})}
\left[1+\frac{\Delta^2+\xi^2(\textbf{k})}{\sqrt{S(\textbf{k})\xi^2(\textbf{k})+h^2\left[\Delta^2+\xi^2(\textbf{k})\right]}}\right]
\end{split}\end{equation}
\begin{equation}\begin{split}
&n_\downarrow(\textbf{k})=1/2-\left[1/2-f_F(\omega_1(\textbf{k})\right]\frac{\xi(\textbf{k})}{2\omega_1(\textbf{k})}
\left[1-\frac{h^2+S(\textbf{k})}{\sqrt{S(\textbf{k})\xi^2(\textbf{k})+h^2\left[\Delta^2+\xi^2(\textbf{k})\right]}}\right]\nonumber\\
&-\left[1/2-f_F(\omega_2(\textbf{k})\right]\frac{\xi(\textbf{k})}{2\omega_2(\textbf{k})}
\left[1+\frac{h^2+S(\textbf{k})}{\sqrt{S(\textbf{k})\xi^2(\textbf{k})+h^2\left[\Delta^2+\xi^2(\textbf{k})\right]}}\right]\nonumber\\
&+\left[1/2-f_F(\omega_1(\textbf{k})\right]\frac{h}{2\omega_1(\textbf{k})}
\left[1-\frac{\Delta^2+\xi^2(\textbf{k})}{\sqrt{S(\textbf{k})\xi^2(\textbf{k})+h^2\left[\Delta^2+\xi^2(\textbf{k})\right]}}\right]\nonumber\\
&+\left[1/2-f_F(\omega_2(\textbf{k})\right[\frac{h}{2\omega_2(\textbf{k})}
\left[1+\frac{\Delta^2+\xi^2(\textbf{k})}{\sqrt{S(\textbf{k})\xi^2(\textbf{k})+h^2\left[\Delta^2+\xi^2(\textbf{k})\right]}}\right]
\end{split}\end{equation}

Another interesting feature of the SOC, is that the pairing field contains both a singlet and
a triplet component. The  singlet $\Phi_{\downarrow\uparrow}(\textbf{k})=-\Phi_{\uparrow\downarrow}(\textbf{k})=<\psi_{\textbf{k}\downarrow}\psi_{-\textbf{k}\uparrow}>$, and  triplet  $\Phi_{\uparrow\uparrow}(\textbf{k})=<\psi_{\textbf{k}\uparrow}\psi_{-\textbf{k}\uparrow}>$, $\Phi_{\downarrow\downarrow}(\textbf{k})=<\psi_{\textbf{k}\downarrow}\psi_{-\textbf{k}\downarrow}>$  amplitudes, obtained by means of the Green's function elements $G_{23}$, $G_{13}$ and $G_{24}$, are:
\begin{equation}\begin{split}
&\Phi_{\downarrow\uparrow}(\textbf{k})=
\frac{\Delta\left[1/2-f_F(\omega_2(\textbf{k})\right]}{2\omega_2(\textbf{k})}
\left[1+\frac{h^2}{\sqrt{S(\textbf{k})\xi^2(\textbf{k})+h^2\left[\Delta^2+\xi^2(\textbf{k})\right]}}\right]\\&+\frac{\Delta\left[1/2-f_F(\omega_1(\textbf{k})\right]}{2\omega_1(\textbf{k})}
\left[1-\frac{h^2}{\sqrt{S(\textbf{k})\xi^2(\textbf{k})+h^2\left[\Delta^2+\xi^2(\textbf{k})\right]}}\right];\nonumber\\
&\Phi_{\uparrow\uparrow}(\textbf{k})=\frac{\Delta J(\textbf{k})\left[h-\xi(\textbf{k})\right]}{2\sqrt{S(\textbf{k})\xi^2(\textbf{k})+h^2\left[\Delta^2+\xi^2(\textbf{k})\right]}}
\left[\frac{1/2-f_F(\omega_2(\textbf{k}))}{\omega_2(\textbf{k})}-\frac{1/2-f_F(\omega_1(\textbf{k}))}{\omega_1(\textbf{k})}\right];\nonumber\\
&\Phi_{\downarrow\downarrow}(\textbf{k})=\frac{\Delta J^*(\textbf{k})\left[h+\xi(\textbf{k})\right]}{2\sqrt{S(\textbf{k})\xi^2(\textbf{k})+h^2\left[\Delta^2+\xi^2(\textbf{k})\right]}}
\left[\frac{1/2-f_F(\omega_2(\textbf{k}))}{\omega_2(\textbf{k})}-\frac{1/2-f_F(\omega_1(\textbf{k}))}{\omega_1(\textbf{k})}\right].
\end{split}\end{equation}

With the help of the pairing amplitudes, one can calculate the condensate fraction $f_c=f_s+f_{tr}$, where $f_s$ and $f_{tr}$ are the singlet and the triplet contributions, correspondingly. At zero temperature, we obtain:
\begin{equation}\begin{split}
&f_s= \frac{2}{N}\sum_{\mathbf{k}}|\Phi_{\downarrow\uparrow}(\textbf{k})|^2\\
&=\frac{\Delta^2}{8N}\sum_{\mathbf{k}}
\left[\frac{1}{\omega_2(\textbf{k})}\left(1+\frac{h^2}{\sqrt{S(\textbf{k})\xi^2(\textbf{k})+h^2\left[\Delta^2+\xi^2(\textbf{k})\right]}}\right)
+
\frac{1}{\omega_1(\textbf{k})}\left(1-\frac{h^2}{\sqrt{S(\textbf{k})\xi^2(\textbf{k})+h^2\left[\Delta^2+\xi^2(\textbf{k})\right]}}\right)\right]^2.
\end{split}\end{equation}
$$
f_{tr}= \frac{1}{N}\sum_{\mathbf{k}}\left(|\Phi_{\uparrow\uparrow}(\textbf{k})|^2+|\Phi_{\downarrow\downarrow}(\textbf{k})|^2\right)
=\frac{\Delta^2}{8N}\sum_{\mathbf{k}}\frac{S(\textbf{k})\left[h^2+\xi^2(\textbf{k})\right]}{S(\textbf{k})\xi^2(\textbf{k})+h^2\left[\Delta^2+\xi^2(\textbf{k})\right]} \left[\frac{1}{\omega_2(\textbf{k})}-\frac{1}{\omega_1(\textbf{k})}\right]^2.
$$

\section{Bethe-Salpeter secular determinant}

The elements of the  $G_{2\times2}(\omega,\mathbf{Q})$, and the
  $D_{8\times 2}(\omega,\mathbf{Q})$ blocks
  are as follows:
\begin{eqnarray}\nonumber
&G_{11}=1+\frac{U}{2}\left(K2233-K1234-K2143+K1144\right),\\\nonumber
&G_{12}= \frac{U}{2}\left(K1414-K2413-K1324+K2323\right),\\\nonumber
&G_{21}=\frac{U}{2}\left(K4141-K4231-K3142+K3232\right),\\\nonumber
&G_{22}=1+\frac{U}{2}\left(K3322-K3412-K4321+K4411\right).\\\nonumber
\end{eqnarray}
Note that $K_{2143}(\omega,\mathbf{Q})$ ($K_{3412}(\omega,\mathbf{Q})$) is complex conjugate of $K_{1234}(\omega,\mathbf{Q})$ ($K_{4321}(\omega,\mathbf{Q})$).
\begin{eqnarray}\nonumber
&D_{11}=-\frac{1}{2}-\frac{U}{2}\left(K3322-K3412\right),
D_{12}= \frac{U}{2}\left(K2323-K1324\right),\\\nonumber
&D_{21}=\frac{U}{2}\left(K1414-K1324\right),
D_{22}=-\frac{1}{2}-\frac{U}{2}\left(K1144-K1234\right).\\\nonumber
&D_{31}=\frac{U}{2}\left(K2322-K2412-K3442+K4414\right).\\\nonumber
&D_{32}=\frac{U}{2}\left(K1224-K1444-K2223+K2443\right).\\\nonumber
&D_{41}=\frac{U}{2}\left(K1321-K1411-K3323+K3414\right).\\\nonumber
&D_{42}=\frac{U}{2}\left(K1114-K1231-K1334+K2333\right).\\\nonumber
&D_{51}=\frac{U}{2}\left(K2411-K2321\right),
D_{52}=\frac{U}{2}\left(K1214-K2213\right).\\\nonumber
&D_{61}=\frac{U}{2}\left(K1412-K1322\right),
D_{62}= \frac{U}{2}\left(K1124+K1232\right),\\\nonumber
&D_{71}=\frac{U}{2}\left(K3432+K4413\right),
D_{72}=\frac{U}{2}\left(K1434-K2433\right).\\\nonumber
&D_{81}=\frac{U}{2}\left(K3414-K3324\right),
D_{82}= \frac{U}{2}\left(K1344-K2343\right),\\\nonumber
\end{eqnarray}
The $8\times 8$ block $C_{8\times8}(\omega,\mathbf{Q})$ is a symmetric block:
\begin{eqnarray}\nonumber
&C_{11}=\frac{1}{2}+\frac{U}{2}K3322, C_{12}=\frac{U}{2}K1324,C_{13}=\frac{U}{2}\left(K3442-K2322\right),\\\nonumber
&C_{14}= \frac{U}{2}\left(K3323-K1321\right),C_{15}=\frac{U}{2}K2321,C_{16}=\frac{U}{2}K1322\\\nonumber
&C_{17}=-\frac{U}{2}K3432,C_{18}=\frac{U}{2}K3324,C_{22}=\frac{1}{2}+\frac{U}{2}K1144,\\\nonumber
&C_{23}=\frac{U}{2}\left(K1444-K1224\right),C_{24}=\frac{U}{2}\left(K1334-K1114\right),\\\nonumber
&C_{25}=-\frac{U}{2}K1214,C_{26}=-\frac{U}{2}K1124,C_{27}=-\frac{U}{2}K1434,\\\nonumber
&C_{28}=-\frac{U}{2}K1344,C_{33}=\frac{U}{2}\left(K2222-K2442-K4224+K4444\right)\\\nonumber
&C_{34}=1+\frac{U}{2}\left(K1221-K1441-K2332+K3443\right),\\\nonumber
&C_{35}=\frac{U}{2}\left(K2212+K2441\right),C_{36}=\frac{U}{2}\left(K1442-K1222\right),\\\nonumber
&C_{37}=\frac{U}{2}\left(K2432-K4434\right),C_{38}=\frac{U}{2}\left(K2342+K3444\right),\\\nonumber
&C_{44}=\frac{U}{2}\left(K1111-K1331-K3113+K3333\right),C_{45}=\frac{U}{2}\left(K1211+K2331\right),\\\nonumber
&C_{46}=\frac{U}{2}\left(K1332-K1112\right),C_{47}=\frac{U}{2}\left(K1431-K3433\right),\\\nonumber
&C_{48}=\frac{U}{2}\left(K1341+K3334\right),C_{55}=\frac{1}{2}-\frac{U}{2}K2211,C_{56}=-\frac{U}{2}K1212,\\\nonumber
&C_{57}=-\frac{U}{2}K2431,C_{58}=-\frac{U}{2}K2341,C_{66}=\frac{1}{2}-\frac{U}{2}K1122,\\\nonumber
&C_{67}=-\frac{U}{2}K1432,C_{68}=-\frac{U}{2}K1342,C_{77}=\frac{1}{2}-\frac{U}{2}K4433,\\\nonumber
&C_{78}=-\frac{U}{2}K3434,C_{88}=\frac{1}{2}-\frac{U}{2}K3344,\\\nonumber
\end{eqnarray}
\end{widetext}
 
\end{document}